\documentstyle[12pt]{article}
\newcommand{\noi}{\noindent}
\newcommand{\eq}{\begin{equation}}
\newcommand{\en}{\end{equation}}
\newcommand{\eqa}{\begin{eqnarray}}
\newcommand{\ena}{\end{eqnarray}}

\newcommand{\aleq}{\mbox{}_{\textstyle \sim}^{\textstyle < }}
\newcommand{\ageq}{\mbox{}_{\textstyle \sim}^{\textstyle > }}

\newcommand{\be}{\begin{equation}}
\newcommand{\ee}{\end{equation}}
\newcommand{\bea}{\begin{eqnarray}}
\newcommand{\eea}{\end{eqnarray}}

\newcommand{\spr}{s^{\prime}}

\newcommand{\bre}{\hfill\break}
\newcommand{\bt}{\beta}
\begin{document}

\hbox{}
\noindent August 1994 \hfill HU Berlin--IEP--94/13
\begin{center}
\vspace*{1.5cm}

\renewcommand{\thefootnote}{\fnsymbol{footnote}}
\setcounter{footnote}{1}

{\LARGE  Compact lattice QED with Wilson fermions}
\footnote{Work supported by the Deutsche
Forschungsgemeinschaft under research grant Mu 932/1-2} \\
\par
\vspace*{1.0cm}
\par
\vspace*{1.5cm}
{\large
A.~Hoferichter $\mbox{}^1$,
V.K.~Mitrjushkin $\mbox{}^1$
\footnote{Permanent adress: Joint Institute for Nuclear 
Research, Dubna, Russia},
M.~M\"uller-Preussker $\mbox{}^1$,
T.~Neuhaus $\mbox{}^2$ and
H.~St\"uben $\mbox{}^3$
}\\
\vspace*{0.7cm}
{\normalsize
$\mbox{}^1$ {\em Humboldt-Universit\"{a}t, Institut f\"{u}r Physik,
10099 Berlin, Germany}\\
$\mbox{}^2$ {\em Universit\"{a}t Bielefeld, Fakult\"{a}t f\"{u}r Physik,
33615 Bielefeld, Germany}\\
$\mbox{}^3$ {\em Konrad-Zuse-Zentrum f\"ur Informationstechnik Berlin,
10711 Berlin, Germany}\\
}  

\vspace*{2cm}
{\bf Abstract}
\end{center}

We study the phase structure and the chiral limit of $4d$ compact
lattice QED with Wilson fermions (both dynamical and quenched).
We use the standard Wilson gauge action and also a modified one
suppressing lattice artifacts.
Different techniques and observables to locate the
chiral limit are discussed.

\newpage
\setcounter{footnote}{1}

\section{Introduction}

QED is commonly recognized to be a very successful quantum field theory.
Nevertheless, despite the excellent agreement found between the
perturbative predictions of the theory and some experimental results,
QED is still a poorly understood theory.
The most unsatisfactory feature of QED is the so--called Landau pole problem.
When a test charge is added to the vacuum of QED,
pair creation around this charge totally screens it and the resulting
charge observed at large distances is zero. That means that QED becomes
'trivial'. Formally we need a cutoff to make it nontrivial.
Lattice (nonperturbative) QED is supposed to resolve this problem.

There are different lattice formulations of QED
(e.g., with compact or non-compact
treatment of the gauge field degrees of freedom
and with staggered or Wilson fermions, respectively)
possibly not belonging to the same class of universality.
If so, on a physical ground one has to decide which version of QED is
realized in nature. Old, i.e. experimentally well--established
physical results should finally be reproduced (at least qualitatively)
from lattice computations.

If we consider QED as arising from a subgroup of a non--abelian
(e.g., grand unified) gauge theory, then we are led to a compact treatment,
a point of view we want to adopt throughout this work.  Concerning the fermion
degrees of freedom we are dealing with Wilson 
fermions \cite{wil,lue}, which -- to our knowledge -- have not 
yet been thoroughly investigated in case of QED.

In a theory with Wilson fermions the chiral symmetry is broken
explicitly. Presumably, it can be restored by fine--tuning
the parameters in the continuum limit. But, for nonzero lattice spacing
we can expect only
a partial symmetry restoration
at some value of the hopping parameter
$\kappa_{c} \equiv \kappa_{c}(\beta )$
by cancellation of
the Wilson mass term with the ordinary mass term
in certain vertex functions at zero momentum \cite{kawa,kasm}.
If so, we can
approach the continuum limit and the chiral symmetry restoration
along the line $\kappa_{c}(\beta )$.
It is another question, whether there the chiral symmetry
becomes explicitly realized or spontaneously broken.

It is worthwhile to note that QED can
also serve as
 a 'toy' model of QCD for the study of, e.g.,
the  chiral transition. Therefore, even the confinement phase of QED is of
  interest, since the fermionic observables should behave
 similar to the case of QCD.

In this work we investigate the phase structure
of 4$d$ lattice QED
with special emphasis on the
chiral limit. We consider the standard Wilson action (WA) and a
modified one (MA) which is derived from WA by introducing
suppression terms for lattice artifacts into the gauge action.
We discuss different techniques to establish the chiral limit.
On the one hand they are based on the convergence rate of the
conjugate gradient algorithm to invert the fermion
matrix, on the other hand they use observables sensitive
to the occurence of zero modes.
Some
preliminary results have been published in \cite{hmmn}.
These techniques are applicable
to lattice QCD studies with Wilson fermions as well.

We concentrate on the study
of gauge invariant observables as a first step, because for
gauge variant quantities such as the photon and fermion correlators
one has to take care of gauge fixing ambiguities
\cite{grib,np,bmmp}.

The organization of the paper is as follows.
In the second section we define both actions and the fermionic
gauge invariant
'order parameters' which we calculated. The third section is devoted
to the discussion of the matrix inversion. The results of our
calculations for the quenched fermions  and
for the dynamical fermions are presented in the fourth
and fifth sections, respectively.
The last section is reserved for the conclusions and the outlook.

\section{Actions and observables}

The standard Wilson lattice action
$ S_{WA}(U, {\bar \psi}, \psi)$ for $4d$ compact $~U(1)~$ gauge 
theory (QED) is

\eq
 S_{WA} = S_{G}(U) + S_{F}(U, {\bar \psi}, \psi) .
                                              \label{wa}
\en

\noi In eq.(\ref{wa}) $~S_{G}(U)~$ is the plaquette
(Wilson) action for the pure gauge  $~U(1)~$ theory

\eq
 S_{G}(U) =
\beta \cdot \sum_{P}
        \,  \bigl( 1 - \cos \theta_{P} \bigr) ~,
                                              \label{wag}
\en

\noi where $~\beta = 1/g^{2}_{bare}~$, and
$~U_{x \mu} = \exp (i \theta_{x \mu} ),
\quad \theta_{x \mu} \in (-\pi, \pi] ~$
are the field variables defined on the links $l = (x,\mu)~$.
Plaquette angles $~\theta_{P}  \equiv \theta_{x;\, \mu \nu}~$
are given by
$~\theta_{x;\, \mu \nu} =
  \theta_{x;\, \mu} + \theta_{x + \hat{\mu};\, \nu}
- \theta_{x + \hat{\nu};\, \mu} - \theta_{x;\, \nu} ~$.

The fermionic part of the action
$S_{F}(U, {\bar \psi}, \psi)~$ is

\eqa
S_{F}
& = & \sum_{f=1}^{N_{f}} \sum_{x,y} \sum_{s,\spr=1}^{4}
{\bar \psi}_{x}^{f,s} {\cal M}^{s \spr}_{xy} \psi_{y}^{f,\spr}
\equiv  {\bar \psi} {\cal M} \psi~,
\nonumber \\ \nonumber \\
{\cal M} & \equiv & \hat{1} - \kappa \cdot Q(U) ,
\nonumber \\ \nonumber \\
Q^{s \spr}_{xy}
& = & \sum_{\mu}
\Bigl[ \delta_{y, x+\hat{\mu}} \cdot ( {\hat 1} - \gamma_{\mu})_{s \spr}
\cdot U_{x \mu} +
\delta_{y, x-\hat{\mu}} \cdot ( {\hat 1} + \gamma_{\mu})_{s \spr} \cdot
U_{x-\hat{\mu}, \mu}^{\dagger} \Bigr]~,
                                              \label{waf}
\ena

\noi where $~{\cal M}~$ is Wilson's fermionic matrix,
$~N_{f}~$ is the number of flavours (we take $~N_{f}~=~2~$
for the case of dynamical fermions),
and $\kappa$ is the hopping parameter.

In our calculations we also
used the modified (compact) action
$S_{MA}$ with Wilson fermions

\eq
 S_{MA} = S_{WA}(U) + \delta S_{G}(U)~,
                                              \label{ma}
\en

\noi where the additional term $~\delta S_{G}~$ is introduced to
suppress some lattice artifacts, i.e., monopoles
and negative plaquettes.
The notion of (DeGrand--Toussaint) monopoles can be introduced
as follows \cite{dgt}~:
the plaquette angle is splitted into two parts
\eq
   \theta_{P} = {\bar \theta}_{P} + 2 \pi n_{P},
   \quad {\bar \theta}_{P}~\in (-\pi,\pi]~,
   \quad n_{P}~=~0, \pm 1, \pm 2,
\en
where $~{\bar \theta}_{P}~$ is associated with the
(gauge-invariant) 'electromagnetic'
flux through the plaquette and $~n_{P}~$ is the number
of Dirac strings passing through it.
The monopole charge within an elementary 3$d$ cube $~c~$ is defined
as the net number of Dirac strings passing through the surface of $~c~$:
\eq
   K_{c}~=~ \frac{1}{2\pi} \sum_{P \in \partial{c}} {\bar \theta}_{P}~=~ 
   -\sum_{P \in \partial{c}} n_{P}
\en
where the correct mutual orientation of the plaquettes has to
be taken into account.
The suppression term in eq. (\ref{ma}) can then be written as
\eq
\delta S_G(U) = \lambda_{K} \cdot \sum_{c} |K_{c}|
                  + \lambda_{P} \cdot \sum_{P}
                  \bigl( 1 - {\rm sign}(\cos \theta_{P}) \bigr)
\en
with $~\lambda_{K}~$ and $~\lambda_{P}~$ as 'chemical potentials'
controlling the suppression rate of the corresponding artifacts.
We have chosen $~(\lambda_{K},\lambda_{P})~=~(\infty,\infty)~$ in order to
obtain maximal suppression of monopoles and negative plaquette values.
Then for the quantized theory the suppression terms can easily be rewritten 
as Kronecker deltas in the measure of the functional integral.

In the case of the pure gauge $~U(1)~$ theory the complete 
suppres\-sion of mo\-no\-po\-les re\-moves the phase 
transition at $~\beta_0 \simeq 1.0~$
\cite{bss,bmm}, and a unique Coulomb phase at positive $~\beta$'s appears 
\cite{bmm}. For staggered fermions no chiral transition is observed as shown
in \cite{qedks}. The additional suppression of negative plaquettes drastically 
improves the overlap with the lowest state, at least, for the case
of the  photon \cite{bmm}. 

It is known, that at least in the strong coupling region for the
standard Wilson action
(in the confinement phase) the ordinary mass term and the Wilson mass term
cancel within the pseudoscalar mass $~m_P~$ at some
$~\kappa = \kappa_c(\beta)~$,
so that quadratic terms in the effective potential vanish for
the pseudoscalar field
, and $\kappa_c=0.25$ at $\beta=0$ \cite{kasm}.

In the weak coupling range (Coulomb phase) perturbative calculations
indicate that the mass of the fermion becomes equal to zero along the
line $\kappa_{c}(\beta )$ (for the free field theory $\kappa_{c}=0.125$)
\cite{kawa}.

Besides the average plaquette $\langle~P~\rangle$
we calculated the following fermionic observables

\eqa
\langle {\bar \psi} \psi \bigr \rangle
& = & \frac{1}{4V} \cdot
\langle \, \mbox{Tr} \Bigl({\cal M}^{-1} \Bigr) \rangle_{G}~,
\\ \nonumber \\
\langle {\bar \psi} \gamma_{5}  \psi \bigr \rangle
& = & \frac{1}{4V} \cdot
\langle \, \mbox{Tr} \Bigl( \gamma_{5} {\cal M}^{-1}\Bigr) \rangle_{G}~,
\\ \nonumber \\
\langle \Pi \rangle
& = & \frac{1}{4V} \cdot \langle \mbox{Tr} \Bigl( {\cal M}^{-1}
\gamma_{5} {\cal M}^{-1} \gamma_{5} \Bigr) \rangle_{G}~,
                                      \label{pionnm}
\ena

\noi where $~\langle ~~ \rangle_{G}~$ stands for averaging over gauge
configurations which in presence of dynamical fermions
includes the fermionic determinant. $~V=L^4~$ is the number of sites.
The advantage of the order parameter in eq.(\ref{pionnm})
-- the 'pion norm' \cite{bkr} -- is that it appears to be a very sensitive
observable in the 'critical' region.
This can be understood by considering the spectral representation
of the fermionic order parameters.
Let $~f_{n} \equiv f_{n}(s,x)~$ be the eigenvectors of
$~{\cal M}~$ with eigenvalues $~\lambda_{n}~$,
and $~g_{n} \equiv g_{n}(s,x)~$ be the eigenvectors of
$~\gamma_{5} {\cal M}~$ with eigenvalues $~\mu_{n}~$ :

\eq
{\cal M} f_{n}
= \lambda_{n} \cdot f_{n}~,
\quad \quad
\gamma_{5} {\cal M} g_{n}
= \mu_{n} \cdot g_{n}~.
\en

\noi Then one can easily obtain a spectral representation
of the fermionic order parameters~:

\eqa
\langle {\bar \psi} \psi \bigr \rangle
& = & \frac{1}{4V} \Bigl \langle~\sum_{n} \frac{1}{ \lambda_{n}}~\Bigr
\rangle_G~,
\qquad
\langle {\bar \psi} \gamma_{5}  \psi \bigr \rangle
= \frac{1}{4V} \Bigl \langle~ \sum_{n} \frac{1}{ \mu_{n}}~\Bigr \rangle_G~,
\nonumber \\ \nonumber \\
\langle \Pi \bigr \rangle
& = & \frac{1}{4V} \Bigl \langle~\sum_{n} \frac{1}{\mu_{n}^2}~\Bigr 
\rangle_G~.
                                 \label{spectral}
\ena

\noi Evidently, an eigenstate of $~{\cal M}~$ with eigenvalue zero is
also eigenstate of $~\gamma_{5} {\cal M}~$. So, the presence of
configurations which belong to zero eigenvalues of $~{\cal M}~$
also gives rise to poles in $~\Pi~$ (for more properties of 
the fermion matrix see, e.g. \cite{iiy}).

It is commonly expected that close to $~\kappa_{c}~$
the minimal eigenvalue $|\lambda_{min }|^2$
of $~{\cal M}^{\dagger} {\cal M}~$ is small and

\eq
|\lambda_{min}|^2 \sim
\Bigl (~ 1 - \frac{\kappa}{\kappa_{c}}~ \Bigr )^2~.
\en

\noi Therefore, the appearance of small eigenvalues $\lambda_i \sim 0$
can be a good indicator for the chiral transition.
For more discussion on this point for the Wilson action see, e.g.,
\cite{blls,uka} and references therein.

In our calculations we have chosen  periodic boundary conditions
for fermions in all four directions. It is worthwhile to notice here
that in the case of quenched fermions the choice between periodic and
antiperiodic boundary conditions is irrelevant.
Indeed, the change of the sign of all, say, time--like links in the
'last' time slice does not change the gauge action, but this corresponds
to the interchange of periodic and antiperiodic boundary conditions
for fermions \cite{nasi}.

In order to extract information about the photon mass
we have measured the photon plaquette--plaquette  correlator at
non--vanishing momenta and for several
values of the coupling parameters (see \cite{bp,bmm}).

\section{ Matrix inversion}

The inversion of the fermionic matrix
$~{\cal M} \equiv \hat{1} - \kappa \cdot Q(U)~$ defined in
eq.(\ref{waf})
is equivalent to solving the matrix equation

\eq
{\cal M} \cdot X = \varphi~.
                                             \label{equa_f}
\en

\noi for any given 'input'--vector $\varphi$.

One of the most popular methods is
the conjugate gradient method (see, e.g., \cite{cg1,cg2}
and references therein).
This is an iterative method to
solve systems of linear equations $D \cdot X = \varphi $,
where $D$ is a hermitian $n \times n$ matrix and
$\varphi$ is an input vector.
The convergence of the conjugate gradient (cg) method
is controlled by the condition number
$~\xi \equiv \lambda_{max}/\lambda_{min}~$,
where $~\lambda_{max}~$ and
$~\lambda_{min}~$ are the maximal and
minimal eigenvalues of $D$
($D = {\cal M}^{\dagger}{\cal M}$ in this case). 
We observed for our standard cg--method that 
at a large enough number of iteration steps
$~N_{cg}~$ the average residue
$~\langle R \rangle \equiv \langle R \rangle (N_{cg})~$ behaves as

\eqa
\langle R \rangle & = & C \cdot \exp (-\alpha \cdot N_{cg} )
\quad \mbox{for} \quad N_{cg} > N_{0},
\nonumber \\
\quad \alpha
& = & \ln \frac{\sqrt{\xi}+1}{\sqrt{\xi}-1}
                                \label{cg1}
\ena

\noi {\it independently of the distribution of eigenvalues}
$\lambda_{i}$ provided $~n~$ is large enough. 
This kind of functional dependence of $~\alpha~$ on $~\xi~$ was analytically
obtained for some other 
gradient methods (e.g. Chebyshev methods) \cite{cg2}.
To check the behaviour (\ref{cg1}) we generated $~D~$ with different
(uniform, gaussian, double--peaked,$\ldots$) distributions
of eigenvalues and given $~\lambda_{min}~$ and
$~\lambda_{{max}}~$.
The components of the input--vector $\varphi$
were chosen randomly with gaussian distributions.

In Fig.1 we show the dependence of the
average (normalized to unity at $N_{cg}=1$) residue
$~\langle R \rangle~$ on $~N_{cg}~$
for a gaussian distribution of eigenvalues $\lambda_i$
of the matrix $D$ in the interval
$~[ \lambda_{min} ; \lambda_{\scriptstyle{max}}]~$
as an illustration.
Broken lines correspond to
the behaviour $~\exp (-\alpha \cdot N_{cg} )~$. One can see
that after some steps $N_{cg} > N_0$ 
eq.(\ref{cg1}) works very well. The behaviour practically
does not depend on the choice of the initial vector $~X_{0}~$:
we checked this for a gaussian distribution of the components $X_{0}(i)$ and
for different 'ordered starts'
(e.g., $X_{0}(1)=1,~X_{0}(i)=0 ~\mbox{at} ~i>1$).
The parameters $N_0$ and $C$ depend on the distribution
of $\lambda_i$.

This observation gives the possibility to use the convergence
of the cg--method to estimate $ ~\kappa_{c}$.
This estimation, however, is biased because of the
dependence of the parameters $~N_0~$ and $~C~$ on the distribution
of eigenvalues.

Obviously, the fermion matrix inversion is the most time consuming step in the
numerical work. Therefore, a maximal improvement of the convergence of the
conjugate gradient algorithm is highly important, in particular near
$~\kappa_c~$.
We have achieved considerable improvement by the use of
preconditioning methods,
e.g., polynomial preconditioning \cite{cg3} plus 'even--odd decomposition'
\cite{dg,dgr}.
Furthermore, the appropriate choice of the start vector $X_0$ allows to speed
up the procedure additionally:

\eq
X_0 = \Bigl[ \hat{1} +
\sum_{l = 1}^{l_0} \kappa^l \cdot Q^l \Bigr] \cdot \varphi~,
                                        \label{hpe}
\en

\noi where $l_0$ is some integer depending on $\kappa$. For smaller
$~\kappa$--values (far enough from $~\kappa_c$)
the hopping parameter expansion (\ref{hpe}) with large
enough $l_0$ already represents a good approximation of the solution
we search for. This approximation is even less time consuming than a combined
application with the cg--method.

\section{Quenched fermions}
\setcounter{footnote}{0}

The quenched approximation of the theory can serve to
provide very useful information about the behaviour of the
fermionic order parameters and to develop the numerical tools.
Its comparison with the full theory is instructive
for the elucidation of the influence of the fermionic determinant.
Moreover, it has the advantage to be less time consuming than the 
full theory. 

As it was already mentioned the phase transition at
$~\beta_0 \simeq 1.0~$ disappears by suppressing
lattice artifacts, and the quenched modified theory has
only one phase
for nonnegative $~\beta$'s :
the Coulomb phase.
Qualitatively, the behaviour of the order parameters we measured
in the modified theory MA at $\bt \ge 0$ turned out to be the same
as in the standard theory WA within the
Coulomb phase $(\bt > \bt_0)$.

Let us now discuss the possibility
to estimate $~\kappa_c(\bt)~$ in the Coulomb phase based on the
convergence of the cg--method.
It follows from the discussion in the previous section
that for the inversion of $~{\cal M}^{\dagger} {\cal M}~$ the number of 
cg--iterations required for the
convergence to some small $R \leq \epsilon$ will behave as
$\langle N^{-1}_{cg} \rangle \sim 1 - \kappa/\kappa_{c}$
as long as  $~N_{cg}~\gg~N_0~$ and $~\kappa \sim \kappa_c~$.
Therefore, the study of the dependence
$~\langle N_{cg}^{-1} \rangle (\kappa)~$ for different
lattice sizes $~L~$ (at fixed
coupling $~\beta~$) can give a rather reliable estimate of
$~\kappa_{c}~$.
Fig.2a represents the dependence of $\langle N^{-1}_{cg} \rangle$
on $~\kappa~$ at $~\beta = 1.1~$ (Coulomb phase) for quenched WA (qWA) for
different $~V=L^4~$ up to $~20^4~$.
By linear extrapolation $\langle N^{-1}_{cg} \rangle \rightarrow 0$
one can obtain reasonable estimates 
of $~\kappa_c~$, which, as we shall see, are
consistent with that obtained by other methods.
The volume
dependence becomes weaker with increasing $~L~$.
With increasing matrix size $~n=4L^4~$ also the variance of $~N_{cg}~$
with respect to different statistical configurations decreases.
An absolutely similar dependence of
$\langle N^{-1}_{cg} \rangle$ on $~\kappa~$ we observed for the
modified theory MA at different values of the gauge coupling.
\footnote{In the confinement phase of WA the number of
required cg--iterations is highly increased and one obtains
even better estimates of $~\kappa_c~$ than
in the Coulomb phase.}

The comparison of the two theories WA and MA
at $~\beta =1.1 ~$ (both in the Coulomb phase) shows
that the difference in the convergence is rather small, and
the estimated values of $~\kappa_c~$ are close to each other. Therefore, the
influence of lattice artifacts in WA is already small at this value
of the coupling. With increasing $~\beta~$ the estimated value
of $~\kappa_c(\beta )~$ shows a
weak dependence on $~\beta~$.

Figs.2b,c show the dependence of $\langle N^{-1}_{cg} \rangle$ on
the inverse lattice size $~L^{-1}~$ at various values of $~\kappa~$ in the
case of the quenched standard Wilson action at $~\beta=1.1~$ (b)
and the quenched modified
action (qMA) at $~\beta=0.8~$ (c) respectively.
For this dependence we again see no qualitative difference 
between Wilson action
in the Coulomb phase and the modified theory and it appears to be
\eq
\langle N^{-1}_{cg} \rangle_{L} = \langle N^{-1}_{cg} \rangle_{\infty}
+ \frac{\mbox{C($\kappa$)}}{L}~.
\en

Time histories of different observables
(e.g., $~\overline{\psi} \psi~, \overline{\psi} \gamma_{5} \psi~,
\Pi~$) can give a signal of approaching $\kappa_c$.
The appearance of sharp peaks in their time histories
at distinct values of $~\beta~$ and $~\kappa~$
testifies the appearance of small eigenvalues
of the fermionic matrix $~{\cal M}~$ and, presumably, could be interpreted
as approaching the chiral limit.

Fig.3  shows the time histories of the pion norm $~\Pi~$
at $~\beta =0 ~$ for the standard Wilson gauge action (confinement phase)
at different $~\kappa$'s.
Strong coupling arguments  predict
a 'critical' value  of $~\kappa~$ at
$~\kappa_{c} = 0.25~$. Indeed, one can see
that below $~\kappa_{c}~$ time histories are rather 'smooth'
and do not show any evidence for peaks. The fluctuations become
more pronounced with increasing $~\kappa~$, and
finally at $~\kappa_{c} \sim 0.25~$ sharp  spikes appear.
It is interesting to note that at $~\kappa > \kappa_c~$ these
spikes do not disappear but instead become even stronger.
With increasing $~\beta~$ (but at $~\beta < \beta_0$)
time histories have a similar structure
with the only difference that the spikes appear at somewhat smaller
values of the Wilson coupling. Thus, e.g.,~
$~\kappa_{c}(\beta=0.8) < \kappa_{c}(\beta=0)~$.
The time histories of
$~\overline{\psi} \psi~$ and $~\overline{\psi} \gamma_{5} \psi~$ look
similar; here the spikes have a smaller order of magnitude in
accordance with the spectral representations.

On the contrary at $~\beta > \beta_{0} ~$ for Wilson  action
(i.e., in the Coulomb phase) the form
of the time histories drastically changes (Fig.4).
We do not find peaks of comparable amplitude ($\sim 10^4$) but,~
nevertheless, in a 'critical' region of $~\kappa~$,
the histories show up stronger fluctuations than at smaller or {\it larger}
values of the Wilson coupling. Taking into account that for
every configuration the pion norm $~\Pi~$ is the
 arithmetic average of $~4 \cdot L^{4}~$
terms corresponding to contributions of $~4 \cdot L^{4}~$
eigenvalues
one can conclude that rather small eigenvalues occured.
A similar picture as in the latter case also holds for the
modified gauge action.

The increase of fluctuations near $~\kappa_c~$ can be best represented
by the statistical variances of the observables discussed,~
especially the variance of $~\Pi~$ should be most sensitive.
The value which we call (renormalized)
variance $~\sigma^2(\Pi )~$ is defined as follows

\eq
\sigma^2(\Pi) = \frac{V}{4^4} \cdot \frac{1}{N}
\sum_{i}^{N} \Bigl( \Pi_{i} - \overline{\Pi} \Bigr)^2~,
\quad
\overline{\Pi} \equiv \frac{1}{N} \sum_{i}^{N} \Pi_{i} ~
\en

\noi where $~\Pi_{i}~$ is the value of
$~\frac{1}{4V} \cdot \sum_{x y}
\mbox{Sp} \Bigl( {\cal M}^{-1}_{~x y} \gamma_{5}
{\cal M}^{-1}_{~y x} \gamma_{5} \Bigr)~$ for the $~i^{th}~$ configuration
and $~N~$ is the number of measurements. ('Sp' denotes the 
trace with respect to Dirac--indices).
This enables us to locate $~\kappa_c(\beta)~$ through the
maximum of $~\sigma^2(\Pi)~$.

Figs.5a,b  show the variance of the pion
norm $~\sigma^2(\Pi )~$ in case of the quenched
Wilson action qWA in the Coulomb phase at $~\beta~=1.1~$ ({\bf a})
and for the quenched modified action qMA
at $~\beta~= 0~$ ({\bf b}).
In both cases there is a clear signal for the chiral
limit or transition at some $~\kappa_c~\sim 0.14~$ ({\bf a})
and $~\kappa_c~\sim 0.135~$ for the case ({\bf b}).\bre
For increasing lattice size $~L~$ the peak at these values of $~\kappa_c~$
becomes more strong and narrow. 

Contrary to this the average pion norm $~\langle \Pi \rangle ~$ itself 
(see Fig.8a--crosses) was found to have only a very weak size dependence.
Moreover, $~\langle \Pi \rangle ~$ shows for qWA in the Coulomb phase and qMA
the same dependence on $~\kappa~$ and it is practically  
independent of $~\beta~$ at fixed $~\kappa~ \le 0.125~$. 

At $~\beta <\beta_0~$ in case of the Wilson action
(confinement phase) the value $~\sigma^2(\Pi )~$
is well defined only {\it below} $~\kappa_c(\beta )~$. As it can
be seen from the time histories for the pion norm the region
$~\kappa > \kappa_c(\beta )~$ ($\beta <\beta_0$) is 'infested' by the
small eigenvalues of the fermionic matrix, so it is practically
impossible to define the average pion norm $~\langle \Pi \rangle ~$ 
and other
fermionic order parameters as well
in this region even with high statistics
($~10.000~$ measurements in our case ).

Based on our quenched data we draw the following phase diagrams in the
($\bt,~\kappa$)--plane~: as already mentioned
the quenched WA has a confinement phase for $\beta < \beta_0$
and a Coulomb phase ($\beta > \beta_0$), and these phases are separated
by a 1st or 2nd order transition line at $~\beta_0~$ (\cite{ln,baig}).
Moreover, we have found a 'critical' line $~\kappa = \kappa_{c}(\beta)~$ which
decreases from $~\kappa_c(\beta=0)= 0.25~$ to
$~\kappa_c(\beta=\infty)=0.125~$.
These lines subdivide the ($\bt,~\kappa$)--plane into four areas which in the
dynamical fermion case are expected to become distinct phases.

The quenched MA has only one Coulomb phase for positive values of $\beta$.
The behaviour of our observables corresponds to that in case of the Coulomb
phase
of qWA. We are also able to locate the line $~\kappa = \kappa_c(\beta)~$
which extends over the whole $~\beta$--range ($~\kappa_c(\beta=0)\sim 0.13~$
and
$~\kappa_c(\beta=\infty)=0.125~$), dividing the phase diagram into 
two possible
phases in the case of full QED.

At $~\bt=0~$ the compact WA and non--compact QED \cite{corn,gerr} agree 
because of the common compact coupling of gauge fields to fermions.
Since the values of $~\kappa_c(\bt)~$ differ for the quenched WA
and quenched MA in the extreme strong coupling limit ($\bt \rightarrow 0$)
our modified theory MA might have different chiral 
properties at strong coupling in comparison with
the non--compact QED with Wilson fermions
(see \cite{corn}).

\section{Dynamical fermions}

For the dynamical fermion case (full QED)
we investigated the phase structure and the chiral transition
for lattice sizes $~4^4,~6^4,~8^4~$ and $~12^4~$.

  We have simulated both actions (standard Wilson and modified)
using the Hybrid
Monte Carlo algorithm \cite{dkpr}.  Throughout our simulations
we have used $N=20$ molecular dynamics steps followed
by an acceptance/rejectance decision at the end of each trajectory.
The length of the trajectories was choosen
for most of our runs to be $N \delta \tau$
equal to unity, resulting into acceptance rates of about $95$ \%
at the considered coupling values. At
a few values of the
couplings we have considered several stepsize values $\delta \tau$
on $4^4$ lattices, where we estimated integrated autocorrelation
times for the plaquette operator from runs of about $10^3$ trajectories.
As far as we can tell, increasing the length
of trajectories up to $N\delta \tau \approx 3$ does not
considerably decrease autocorrelation times, while a further increase
leads to small acceptance rates and to a deterioration of the achieved 
MC statistics.

Concerning the simulations with the modified action (\ref{ma}) we have chosen
a unique suppression term $\delta S_G(U)$ in the form
\eq
\delta S_G(U) = A \cdot \sum_{P} \,
\Bigl( 1 + \tanh ( B \cos \theta_{P} ) \Bigr)~,
                                              \label{asupr}
\en
since the Kronecker deltas in the functional integral measure cannot
be realized in the molecular dynamics part of the
Hybrid Monte Carlo procedure
in a straightforward way.
\noi With an appropriate choice of the parameters $~A~$ and $~B~(B < 0)~$
the term (\ref{asupr}) leads to a suppression of negative plaquette values.
The parameters in eq. (\ref{asupr}) were chosen to be $A=10$ and $B=-10$.
We have monitored
the distribution functions of local plaquette values~:
In Fig.6 we display the distribution
function of the local plaquette operator at the same values of couplings
with (crosses) and without (dots) suppression term as an example.
We see, that this choice of parameters leads to an almost
complete suppression of negative plaquette values.
Moreover, we made sure that in a simulation with a
cold start monopoles never occur.

In the simulations with suppression term
the length of the trajectories was choosen to be almost a factor of
$10$ smaller in order to achieve reasonable acceptance rates of
about $90$ \%.  It appears as if the constraint on the action
reduces the available phase space considerably and allows only
for smaller moves in the Hybrid Monte Carlo algorithm.

\vspace{0.25cm}

Now let us turn to the question how the phase diagrams of
the quenched theories given at the end of the previous section
are changed when we allow for dynamical fermions.

For the Wilson action WA we expect to have
at least one additional 'upper left' and one 'upper right' phase
which for simplicity
we will refer to as the '4rd' (upper
left) and '3rd' (upper right) phases (see Fig10a).
We do not consider the question of possible additional phases 
at larger $\kappa $--values \cite{kawa}.

At $~\beta \simeq 1.0~$ and $~\kappa < \kappa_1 \sim 0.18~$
we observe the same
phase transition as in the quenched theory.
Thermal cycles and
time histories for the average plaquette show a clear signal
of metastability at $~\beta$'s close to $~\beta_0 ~$
and at different  $~\kappa < \kappa_1~$.
As well as in the quenched case this transition is driven
by monopole condensation in the confinement phase.
However, there is no such signal
for $~ {\bar \psi} \psi ~$, which remains approximately constant
at fixed $~\kappa~$ and varying $~\beta~$
around $~\beta_0~$ as measured at $~\kappa~=~0.06,~0.08,~0.10~$.
Therefore, at the moment we cannot conclude that
there is a first order phase transition
in this part of the phase diagram in disagreement with \cite{hege}.
Probably the metastable states are due to the monopole loops
wrapping around the torus.

At $~\beta < \beta_0~$ thermal cycles with respect
to varying $~\kappa~$ showed a typical
hysteresis behaviour and a metastability
for both~:
the plaquette and $~{\bar \psi} \psi ~$.
In Fig.7a we display the hysteresis loop
for $~{\bar \psi} \psi ~$ at $~\beta =0.8~$.
Furthermore, the abelian monopole density strongly decreases
with rising $~\kappa~$ (Fig.7b).
In both of the figures  the full (open) symbols correspond to
the runs with decreasing (increasing) $~\kappa~$.

For the standard compact gauge action case we therefore expect
a 'horizontal' first order transition line
from $~(\beta,~\kappa_c)~=~(\beta_1 ~\aleq~ \beta_0,~\kappa_1 ~\sim ~0.18)~$ to
some $~(\beta,~\kappa_c)~=~(\beta_2 ~\ageq ~0.15,~\kappa_2 ~\ageq~ \kappa_1)~$,
 the latter point assumed to be the 'right--lower corner' of the 4th phase 
(cf. discussion below).
This transition is also
driven by monopole condensation in the confinement phase.
Yet we do not know what kind of transition line the phase boundary between
the 4th and the confinement phase is, i.e. the 'horizontal' line 
continued from
$~(\beta,~\kappa_c)~=~(\beta_2,~\kappa_2)~$ to
the point $~(\beta,~\kappa_c)~=~(0.,0.25)~$ already known from
strong coupling expansion.

\vspace{0.25cm}
The study of the (gauge invariant) photon plaquette--plaquette
correlator
$~\Gamma_{\gamma}(\tau )~$ can serve to distinguish between the
different phases.
In the confinement phase of WA
$~\Gamma_{\gamma}(\tau )~$ decays very quickly with increasing
$~\tau~$, so that
for the photon mass $~m_{\gamma}~$
the conclusion $~m_{\gamma} > 0~$ can be drawn.
In the Coulomb phase
the behaviour of the correlator is consistent with the
expected zero mass behaviour.
In the 3rd phase our data also indicate the
existence of a massless photon, but still more effort is needed
to draw the final
conclusion about the photon mass in the two 'upper' phases.
\footnote{In our preliminary publication \cite{hmmn} we
conjectured a tachyonic 
behaviour, i.e. $~m^2_{\gamma} <0~$, in the 3rd phase.}

\vspace{0.25cm}

In the following we want to discuss the chiral limit or transition from the
Coulomb phase to the higher--$\kappa$ phase(s).

As in the case of the quenched theory
the study of the time histories of different observables,
in particular,
$~\Pi~$, can give a signal of approaching $\kappa_c$.
However, in the case of the theory with dynamical fermions
the fermionic determinant
strongly suppresses
the small
eigenvalues, and there are no such
sharp peaks in the time histories as in the case of the
quenched theory.
Nevertheless, the study of time histories shows that
by approaching $~\kappa_c(\beta )~$
they become much more noisy than far away from it,
testifying the appearance of small eigenvalues
of the fermionic matrix $~{\cal M}~$ (cf. \cite{bkr}).

Again the statistical variance should be a suitable parameter
to express the fluctuations of the observables.

The behaviour of the pion norm $~\langle \Pi \rangle~$ and
its variance in the full QED is similar to the behaviour of
these quantities in the quenched theory.

Fig.8a shows the dependence of $~\langle \Pi \rangle~$ on
$~\kappa~$ at $~\beta =1.1~$ for the standard Wilson action.
One can see that the pion norm has a pronounced maximum whose
position and value practically does not depend on the lattice
size at large enough $~L~$ ($L=8$ and $12$ in our case).
At smaller values of $~L~$ there is still a finite volume dependence.
For comparison we show the data points (crosses) for $~\langle\Pi\rangle~$
at the same values of $~\beta~$ and $~\kappa~$, but for a quenched simulation
on a $~8^4~$ lattice -- this illustrates the influence of the
vacuum polarization effects caused by dynamical fermions.

For the variance $~\sigma^2(\Pi )~$ of the
pion norm (Fig. 8b) the signal in the full QED case is not
as strong as in the quenched theory (smaller by a factor of $~\sim$ 6),
because of the effect of the fermionic determinant.
Nevertheless, the maximum of the variance $~\sigma^2(\Pi )~$ of the
pion norm shows a steady growth with increasing $~L~$.
(We expect the errorbars shown in the figure to be somewhat
underestimated (at most $\sim 15\%$) since we have
seen strong autocorrelations in our data for
$~\sigma^2(\Pi )~$, especially around $~\kappa_c~$.)
We also observe a shift of the maximum of $~\sigma^2(\Pi )~$ 'to the left'
the more $~L~$ is increased. Therefore, the variance of the pion norm seems to
exhibit the characteristic features of a finite size scaling behaviour
of a singular point on finite volume systems.

We are led to the conclusion, that a chiral phase transition
in the thermodynamic sense really exists. We might speculate that the
corresponding singularity appears to be rather
'weak', i.e., other operators and their fluctuations, like the 
fermionic condensates and  the plaquette operator 
do not show a singular behaviour, as far as we can tell for the
given lattice sizes. 
As the variance of the pion norm itself can
be interpreted as a higher derivative operator we speculate that the
singularity may be viewed as a higher order phase transition.

Based on our observations we conclude that the variance of the 
pion norm $~\sigma^2(\Pi )~$ is a convenient 'order parameter'
for determining the position of the chiral transition
in the Coulomb phase. On the other hand
the average pion norm itself, its time histories and 
$~\langle N_{cg}^{-1} \rangle~$ can provide estimates of $\kappa_c$.

Turning to the investigation of the phase structure
of the theory MA (eq.(\ref{ma})) the situation drastically changes
in comparison with the standard theory WA.
Similarly to the quenched theory in the full QED with artifacts
suppressed we see only one phase at $~\kappa < \kappa_c(\beta )~$,
and  only one unique phase just above the line $~\kappa_c(\beta )~$.
Moreover, we do not observe metastable states when crossing the
'horizontal' line, and no hysteresis shows up in thermal cycles.
The average plaquette and the fermionic condensate, respectively,
turn out to behave smoothly with varying $~\kappa~$ at fixed $~\beta~$.
This has been checked for $~\beta~=~0~$ and $~\beta~=~1.1~$.
The corresponding curves look very similar.
So, there is no sign for a first order transition.

We have also explored the small $\beta$--region of the phase
diagram above the chiral phase transition $\kappa_c(\beta)$
in the standard Wilson action formulation.
As it was already noted strong coupling arguments suggest
for $\beta=0$ and for $\kappa$--values
above $\kappa_c(\beta=0)=0.25$ the existence
of zero eigenvalues of the fermion matrix.
It is a valid question how this singularity is approached
in the dynamical fermion simulation
for positive values of $\beta$, and for $\kappa > \kappa_c(\bt)$, i.e.  
above the chiral phase transition line.
In particular one might ask whether this singularity of the theory
extents to finite $\beta$--values and whether a corresponding
unphysical phase of the theory exists.

 We have simulated the approach to small $\beta$-values at a fixed
Wilson coupling value $\kappa=0.3 > \kappa_c(\bt)$ on a $4^4$ lattice
for various $\beta$-values in runs of $1000$ measurements, i.e. 
4000 trajectories.

In this region of the phase diagram  we encountered acceptance rate
problems in the Hybrid Monte Carlo algorithm. 
For this reason we have decreased the stepsize $\delta \tau$ to 0.025
while keeping the trajectory length fixed at unity -- the resulting
acceptance rate turned out to be about 84 \%.

With decreasing $~\beta~$ we observe a rapid increase
of the number of the conjugate gradient steps  $~N_{cg}~$
needed for the fermion matrix inversion (Fig.9a).
A linear extrapolation of the average inverse  number
$~\langle N^{-1}_{cg}\rangle ~\rightarrow 0$  of conjugate gradient steps
(dotted line in Fig.9a)
can be consistent with a singular point around $~\beta \sim 0.15$ .\bre
In Fig.9b we display the time evolution of the pion
norm  $~\Pi~$ at decreasing values of $~\beta~$.
The more $~\beta~$ is lowered the more sharp spikes with increasing amplitude
are observed in the time evolution of the
operator, similar to that in the quenched theory
(compare with Fig.3).
The appearance of such spikes makes it
practically impossible
to calculate a statistically reliable average value.\bre
We therefore
conclude that the point  $~\beta \sim  0.15~$ at $~\kappa =0.3~$
is part of a singular line $~\bt = \bt(\kappa), \kappa > 0.25~$ which
separates the 4th from the 3rd phase.

The 4th phase could be expected to be a parity--violating
phase (see \cite{aoki1}).
In case of QCD the existence of a parity violating--phase 
for $~\kappa > \kappa_c(\bt)~$
was numerically based on the behaviour of
the fluctuation of the pseudoscalar density
$~\langle \bar{\psi}\gamma_5\psi\rangle~$ \cite{aoki}.
We studied the distribution of 
$~\bar{\psi}\gamma_5\psi~$ at $~(\beta=0~;~\kappa < \kappa_c)~$ 
and $~(\beta=0~;~\kappa > \kappa_c)~$ in order
to find a signal for
parity--violation in the quenched approximation 
of the standard Wilson action on a $4^4$  lattice.
The distribution width increases roughly by a
factor of 6 while crossing $~\kappa_c~$
from below. This is due to the appearance of
small eigenvalues of $~\cal{M}~$ for
$~\kappa > \kappa_c~$ leading to large fluctuations
of $~\bar{\psi}\gamma_5\psi~$ seen as spikes in the time history.
The
shape of the distribution does not change, especially we do not find
any double--peak structure for $~\kappa~ > \kappa_c~$.
One has to repeat this analysis on a larger lattice to control the finite
size effects. Preliminary our data do not show parity--violation
for the considered $~(\beta,~\kappa)$--values.

The phase diagrams for both theories (with standard gauge action and
with modified action) are shown in Figs.10a,b.
For the standard action WA (Fig.10a) we show four different phases at
$~\kappa \aleq 0.3~$, whereas for the modified theory MA only
two phases survive in the same $~\kappa$--range (Fig10b).

\vspace{0.25cm}

\section{Conclusions and Outlook}

Now let us summarize our results.

\begin{itemize}

\item We have studied the phase structure of two theories with
Wilson fermions and compact action with $U(1)$ symmetry :
standard Wilson theory and a modified one with lattice artifacts
suppressed. Phase diagrams of both theories were shown.

\item For the standard Wilson theory there is a 'horizontal'
line $~\kappa_c(\beta )~$ from a $(\beta_1; \kappa_1)$
to $(\infty; \frac{1}{8})$ which separates the Coulomb phase from the
3rd phase and which corresponds to the {\it chiral transition}
in this theory.
Presumably, a higher order phase transition occurs on this
line.
At smaller values of $~\beta~$ ($\beta < \beta_1$) on the continuation
of this line we find a presumably first order chiral phase transition.

\item When suppressing the lattice artifacts the latter 1st order
phase transition disappears.

\item  Nevertheless, after sup\-pres\-sing the lattice arti\-facts a chi\-ral
tran\-si\-tion (i.e., ap\-pearance of near--to--zero eigenvalues
of $~{\cal M}~$) remains for all $~\beta~$ values
we considered. The statistical variance of the pion norm gives us a
pronounced signal for both (quenched and dynamical) cases.
Its location turned out to be in reasonable agreement with that obtained
from the cg--method (measuring $~\langle N^{-1}_{cg} \rangle~$).

\item  The behaviour of all considered gauge--invariant observables
{\it along} the critical line $~\kappa_c(\beta)~$ in the Coulomb phase
is smooth.
Thus, for the modified action,
we have no indication for a (tri--) critical point on this
line, in contrast to lattice QED with staggered fermions and the non-compact
gauge action.

\item

The calculation of the (gauge--variant) fermionic propagator
needs a gauge--fixing procedure which can appear to be
rather nontrivial because of gauge fixing ambiguities \cite{grib}.
In a recent paper \cite{np},~ 
some of the gauge copies were shown to
produce a photon propagator with a decay behaviour
inconsistent with the expected zero mass behaviour in the Coulomb phase.
In Ref. \cite{bmmp} it was proven that the "bad" gauge copies
are due to the existence of pairs of {\it Dirac sheets} as gauge artifacts.
For matter fields to be taken into account this problem
requires additional studies before one can reliably compute fermionic
correlators.

\item

As a further step we plan to investigate the spectra
of both theories and to
determine renormalized quantities such as the
renormalized coupling.

\end{itemize}

\noi {\large \bf Acknowledgements}
\par\bigskip\noi

\noi We would like to acknowledge useful discussions with 
E. Laermann, P.~Rakow, G.~Schierholz, R.~Sommer and U.-J.~Wiese. 

\newpage

\noi {\large {\bf  Figure captions.}}

\vspace{0.25cm}
\noi {\bf Fig.1} The average residue $\langle R \rangle$ as a
function of $~N_{cg}~$ for gaussian distribution of eigenvalues
$\lambda_i$ of the hermitian matrix $D$ in the interval
$~[\lambda_{min};\lambda_{max}]$.
Broken lines correspond to the behaviour
$~\exp (-\alpha \cdot N_{cg} )~$ where $\alpha$ is defined in eq.(\ref{cg1}).

\vspace{0.2cm}
\noi {\bf Fig.2a} The dependence of
$~\langle N^{-1}_{cg} \rangle~$ on $~\kappa~$ at $~\beta = 1.1~$
for the quenched Wilson action (qWA) and different lattice sizes $L$.
The broken line has been added to guide the eye.

\vspace{0.2cm}
\noi {\bf Figs.2b,c} $~\langle N^{-1}_{cg} \rangle~$ as a function
of the inverse lattice size $L^{-1}$ at different values of
$~\kappa~$ for~
({\bf b})~qWA at $~\beta = 1.1~$ and~
({\bf c})~the quenched modified action (qMA) at $~\beta = 0.8~$ .
Broken lines are to guide the eye.

\vspace{0.2cm}
\noi {\bf Fig.3} Time histories of $\Pi$ at different $~\kappa~$
at $~\beta =0~$ for qWA.

\vspace{0.2cm}
\noi {\bf Fig.4} Time histories of $\Pi$ at different $~\kappa~$
at $~\beta =1.1~$  for qWA.

\vspace{0.2cm}
\noi {\bf Figs.5a,b} $~\sigma^2 (\Pi ) ~$
as a function of $~\kappa~$ for qWA at
$~\beta = 1.1~$ ({\bf a}) and the qMA at
$~\beta = 0~$ ({\bf b}). The broken lines indicate the maxima
of
$~\sigma^2 (\Pi ) ~$.

\vspace{.25cm}
\noi {\bf Fig.6} Distribution functions of the plaquette operator
without (dots) and with (crosses) suppression term (\ref{asupr}) at
$~\beta=0.8~,~\kappa=0.1$ on a $6^4$  lattice.

\vspace{0.2cm}
\noi {\bf Figs.7a,b} The thermal cycle for $~{\bar \psi} \psi~$ ({\bf a})
and for the  monopole density $~\rho_{mon.} ~$ ({\bf b})
with respect to $~\kappa~$ at $~\beta = 0.8~$
for  WA with dynamical fermions on a $6^4$ lattice.

\vspace{0.2cm}
\noi {\bf Figs.8a,b} Average pion norm $~\langle \Pi \rangle~$
({\bf a}) and $~\sigma^2 (\Pi ) ~$ ({\bf b})
as a function of $~\kappa~$ for WA at
$~\beta = 1.1~$.
Lattice size is $~4^4~$ (circles), $~6^4~$ (squares) ,
$~8^4~$ (diamonds) and $~12^4~$ (triangles).
For comparison the data from a quenched simulation on a $~8^4~$ lattice is
shown (crosses) in Fig.8a.

\vspace{0.2cm}
\noi {\bf Figs.9a,b} $~\langle N^{-1}_{cg} \rangle~$ as a 
function of $~\beta~$
at $~\kappa =0.3~$ ({\bf a}) and time histories of $~\Pi~$
at $~\kappa =0.3~$ and and different $~\beta ~$ ({\bf b})
for WA on a $~4^4~$ lattice.

\vspace{0.2cm}
\noi {\bf Figs.10a,b} Phase diagram in the 
$(\beta,~\kappa )$--plane
for dynamical fermions for  the standard Wilson action WA ({\bf a})
and for the modified action MA ({\bf b}).


\vspace{1cm}
\begin{thebibliography}{99}

\newcommand{\prd}[1]{Phys.~Rev.~{\bf D#1}\ }
\newcommand{\plb}[1]{Phys.~Lett.~{\bf #1B}\ }
\newcommand{\npb}[1]{Nucl.~Phys.~{\bf B#1}\ }
\newcommand{\prl}[1]{Phys.~Rev.~Lett.~{\bf #1}\ }
\newcommand{\pr}[1]{Phys.~Rep.~{\bf #1}\ }
\newcommand{\ap}[1]{Ann.~Phys.~{\bf #1}\ }
\newcommand{\cmp}[1]{Commun.~Math.~Phys.~{\bf #1}}
\newcommand{\rmp}[1]{Rev.~Mod.~Phys.~{\bf #1}}
\newcommand{\ptp}[1]{Prog.~Theor.~Phys.~{\bf #1}}
%

\bibitem{wil}   K. Wilson, \prd{10} (1974) 2445;
                in New phenomena in subnuclear physics, ed. A. Zichichi
                (Plenum, New York, 1977)
\bibitem{lue}   M. L\"uscher, Comm. Math. Phys. 54 (1977) 283
\bibitem{kawa}  N. Kawamoto, Nucl. Phys. {\bf B190} (1981) 617
\bibitem{kasm}  N. Kawamoto and J. Smit, Nucl. Phys. {\bf B192} (1981) 100
\bibitem{hmmn}   A. Hoferichter, V.K. Mitrjushkin, M. M\"uller--Preussker
                and Th. Neuhaus, 
                Nucl.Phys. {\bf B} (Proc. Suppl.) 34 (1994) 537
\bibitem{grib}  V.N.~Gribov, \npb{139} (1978) 1
\bibitem{np}    A. Nakamura and M. Plewnia, \plb{255} (1991) 274
\bibitem{bmmp}  V.G. Bornyakov, V.K. Mitrjushkin, M. M\"uller-Preussker
                and F.Pahl,  \\ Phys. Lett. {\bf B317} (1993) 596
\bibitem{dgt}   T.A. DeGrand and D. Toussaint,
                Phys. Rev. {\bf D22} (1980) 2478
\bibitem{bss}   J. S. Barber, R. E. Shrock and R. Schrader,
                ~Phys. Lett. {\bf B152} (1985) 221 \\
                J. S. Barber and R. E. Shrock,
                \npb{257} [FS 14] (1985) 515
\bibitem{bmm}   V.G. Bornyakov, V.K. Mitrjushkin and M. M\"uller--Preussker,\\
                Nucl. Phys. {\bf B} (Proc. Suppl.) 30 (1993) 587
\bibitem{qedks} A. Hoferichter, V.K. Mitrjushkin and M. M\"uller--Preussker,\\
                HU Berlin--IEP--94/11, hep-lat/9407009
\bibitem{bkr}   K.~Bitar, A.~D.~Kennedy and P.~Rossi,
                Phys. Lett. {\bf B234} (1990) 333
\bibitem{iiy}   S. Itoh, Y. Iwasaki and T. Yoshi\'{e},
                Phys. Rev. {\bf D36} (1987) 527
\bibitem{blls}  I. Barbour, E. Laermann, Th. Lippert and
                K. Schilling, \\ Phys. Rev. {\bf D46} (1992) 3618
\bibitem{uka}   A. Ukawa, CERN-TH-5245/88 (1988)
\bibitem{nasi}  A. Nakamura and R. Sinclair, \plb{243} (1990) 396
\bibitem{bp}    B.~Berg and C.~Panagiotakopoulos, Phys. Rev. Lett. {\bf 52}
		    (1984) 94
\bibitem{cg1}   M. R. Hestenes and E. Stiefel, Journal of Research of the NBS
                {\bf 49} (1952) 409
\bibitem{cg2}   M.~Engeli et al., {\it  Mitteilungen aus dem Institut f\"{u}r
                angewandte Mathematik} (Birkh\"{a}user Verlag,
                Berlin, 1959), Vol. 8, p. 24
\bibitem{cg3}  A. Nakamura, G. Feuer, H.C. Hege and V. Linke, \\
               Comp. Phys. Comm. {\bf 51} (1988) 301
\bibitem{dg}   Th. A. DeGrand,
               Comp. Phys. Comm. {\bf 52} (1988) 161
\bibitem{dgr}  Th. A. DeGrand and P. Rossi,
               Comp. Phys. Comm. {\bf 60} (1990) 211
\bibitem{corn}  A. Cornelius, thesis Hamburg University (1991), unpublished
\bibitem{gerr}  G. Schierholz,
                Nucl. Phys. {\bf B} (Proc. Suppl.) 20 (1991) 623
\bibitem{dkpr}  S. Duane, A. D. Kennedy, B. J. Pendleton and
                D. Roweth,  \\ Phys. Lett. {\bf B195} (1987) 216
\bibitem{hege}  C. Hege and A. Nakamura,
                Nucl. Phys. {\bf B} (Proc. Suppl.) 9 (1989) 114
\bibitem{ln}    C.B. Lang and T. Neuhaus,
                Nucl. Phys. {\bf B} (Proc. Suppl.) 34 (1994) 543; \\
		    C.B. Lang and T. Neuhaus, BI--TP 94/37, hep-lat/9407005
\bibitem{baig}  M. Baig and H. Fort, UAB--FT--338
\bibitem{aoki1}  S.~Aoki, Phys. Rev. {\bf D30} (1984) 2653
\bibitem{aoki}  S.~Aoki, Phys. Lett. {\bf B190} (1987) 140
\end{thebibliography}
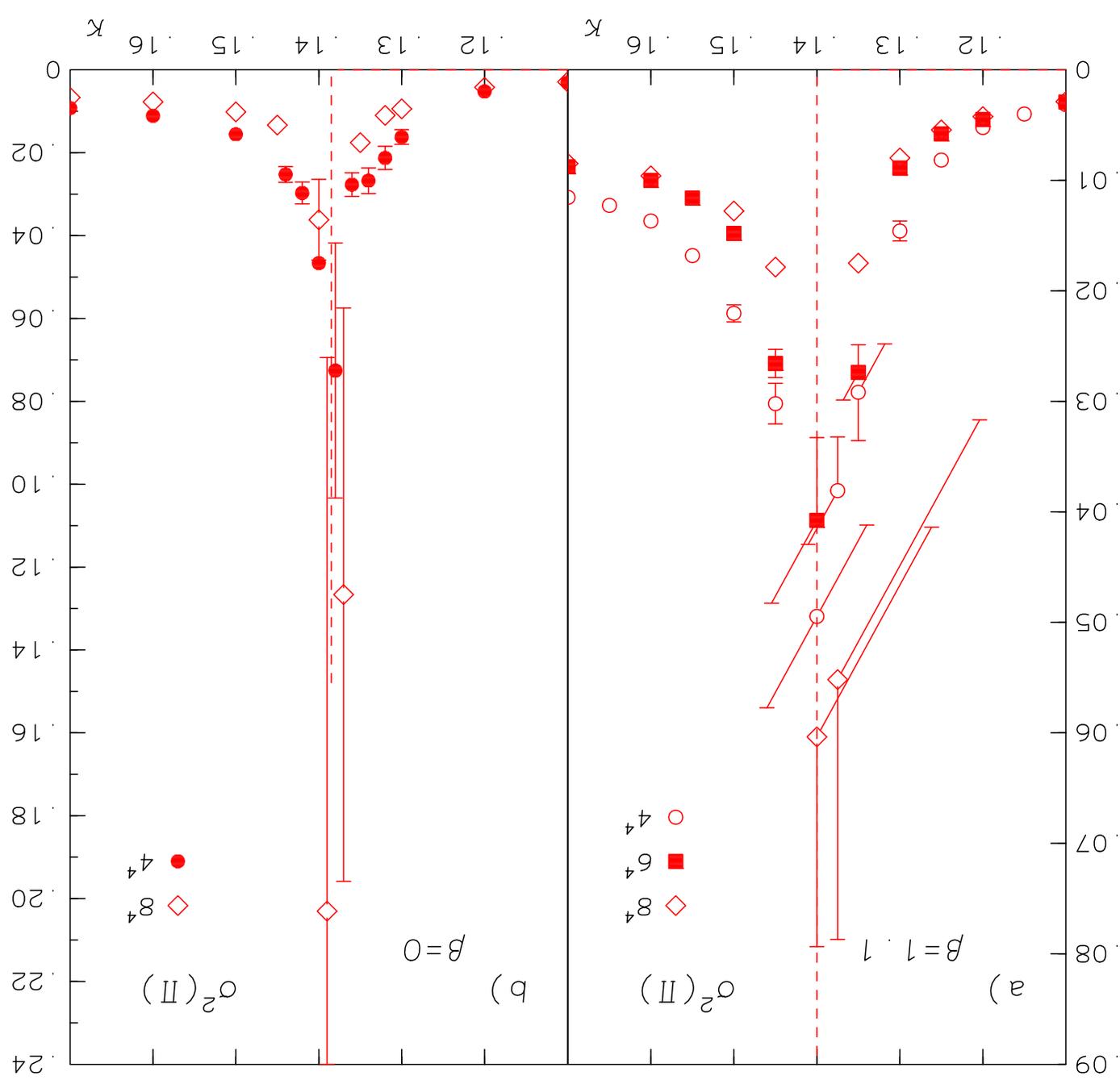
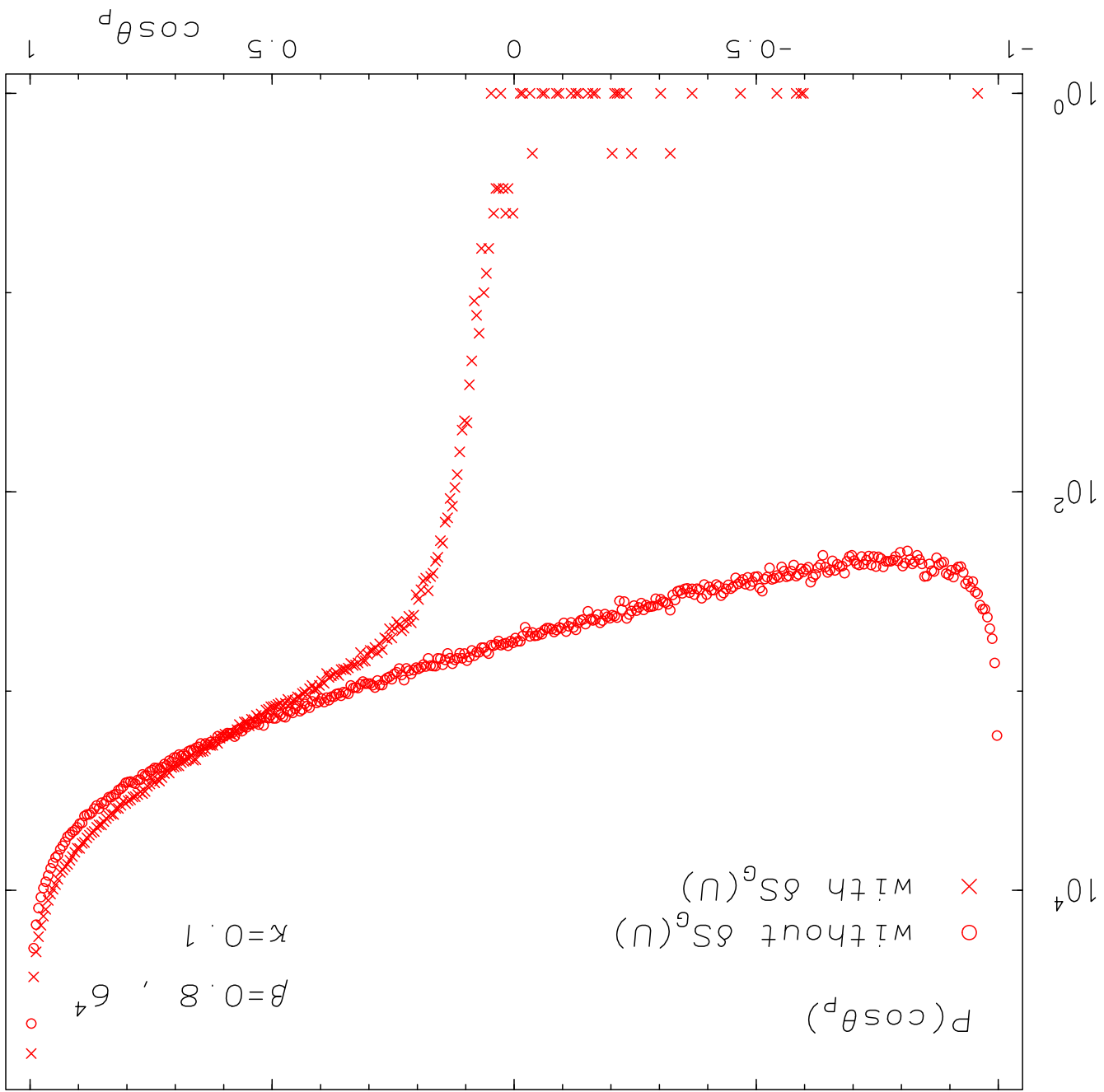
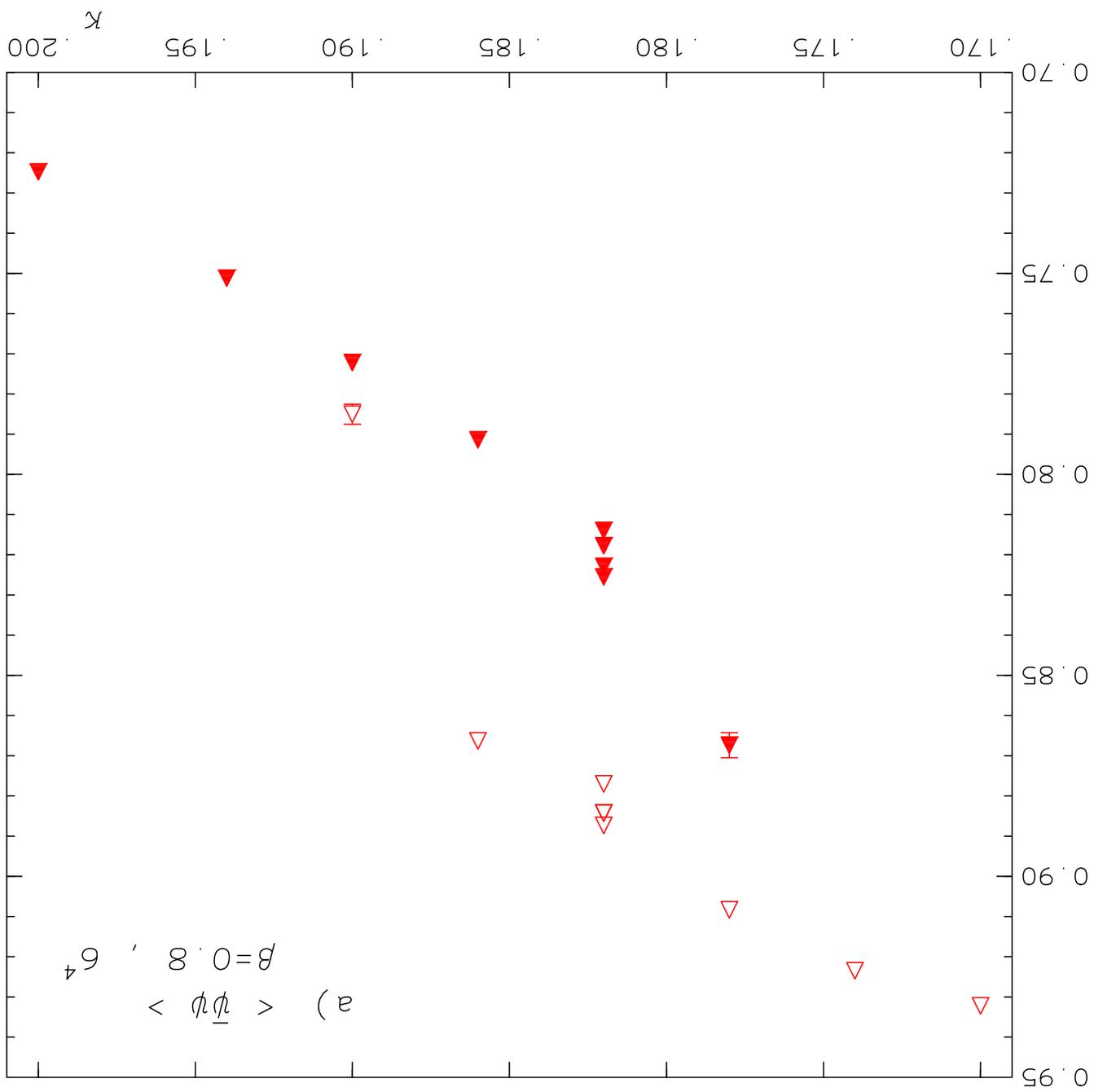
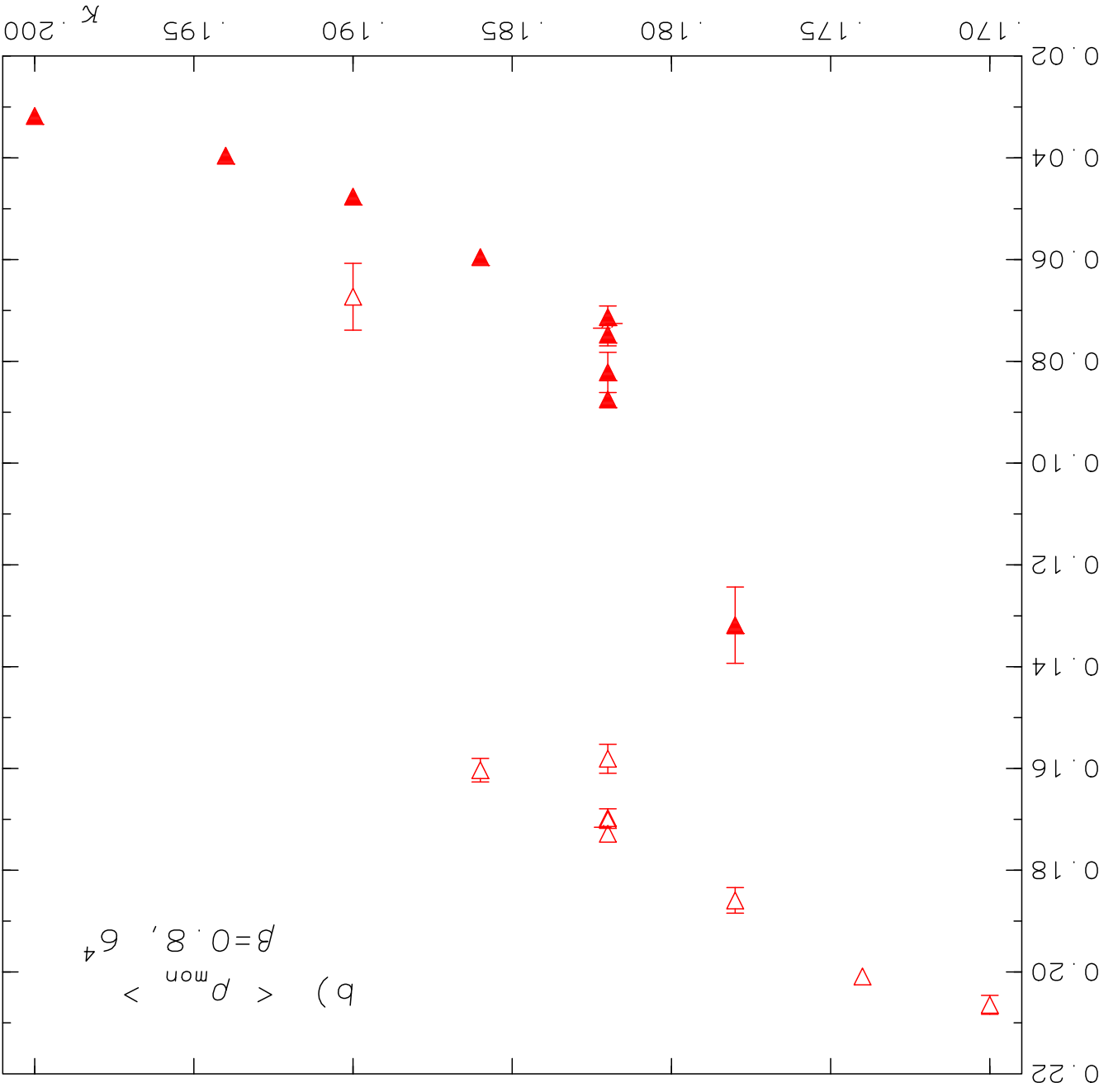
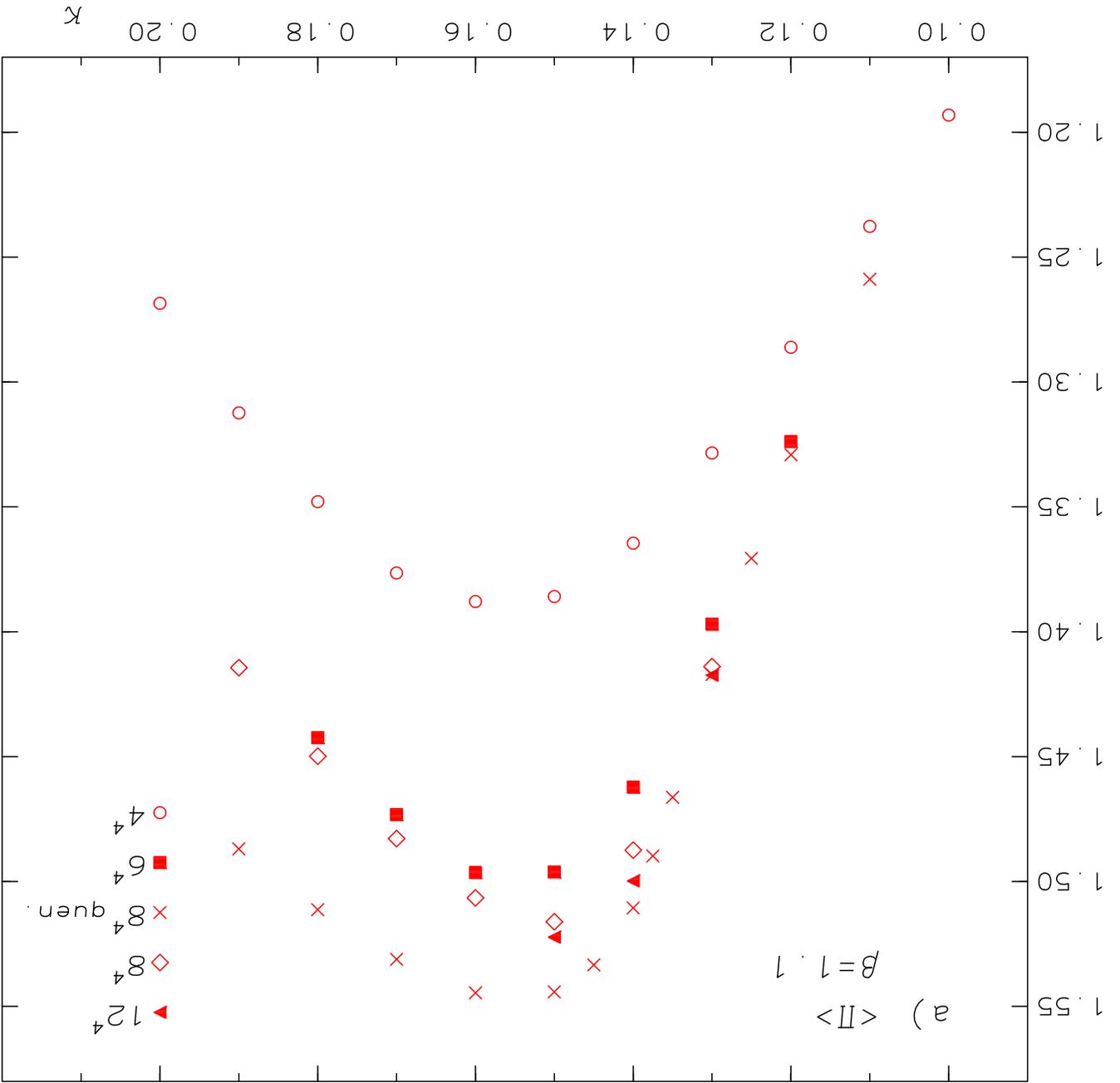
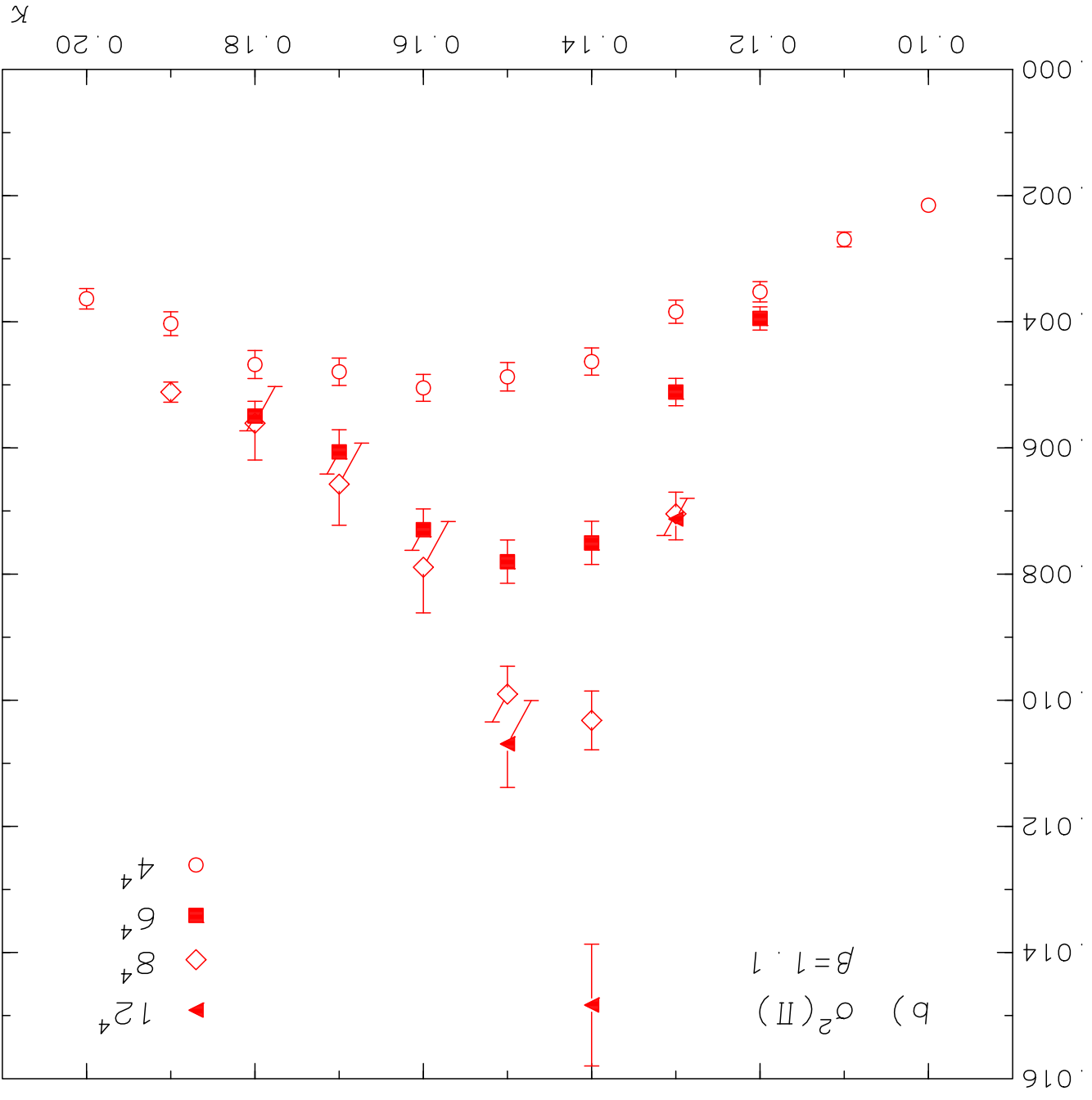
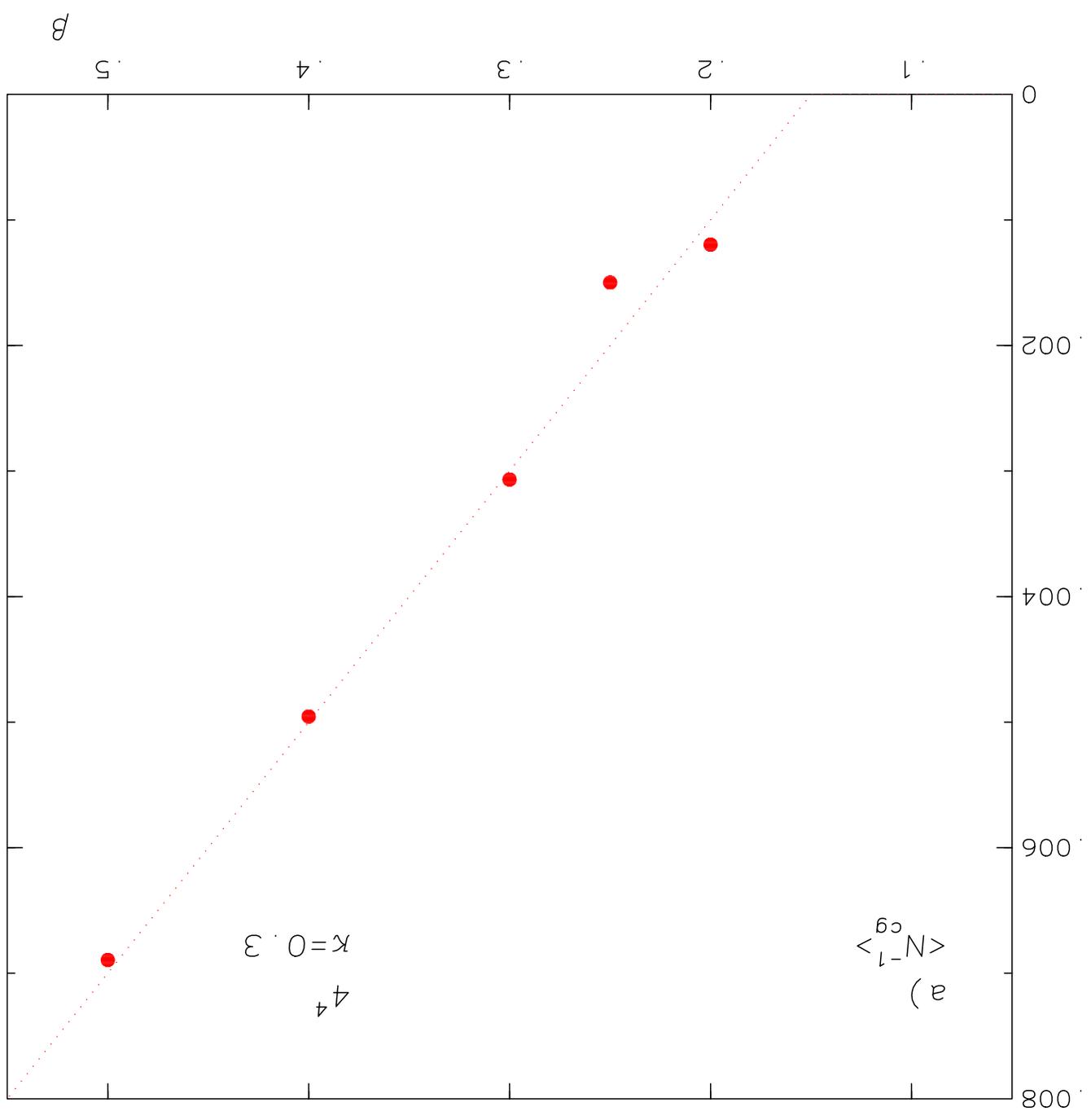
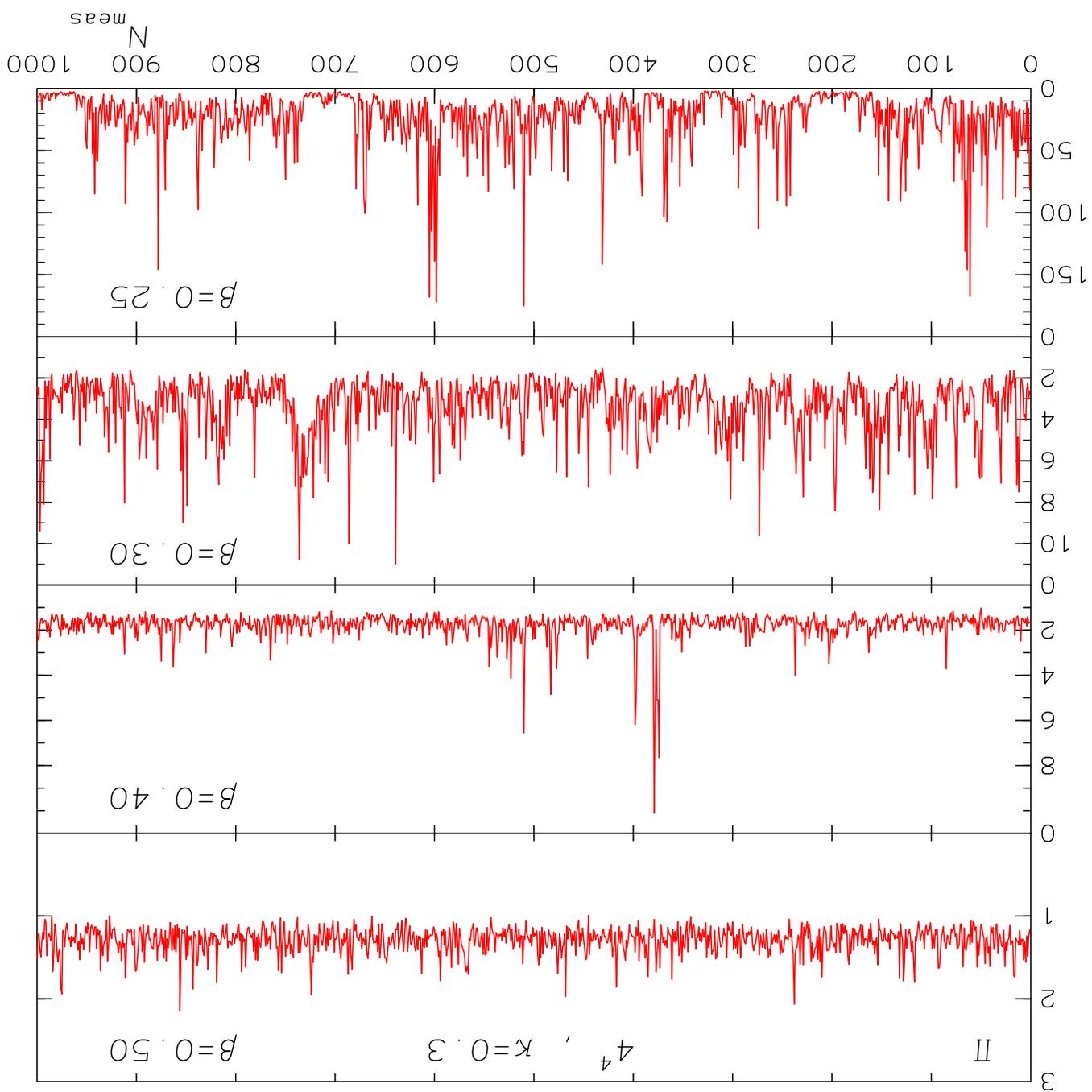
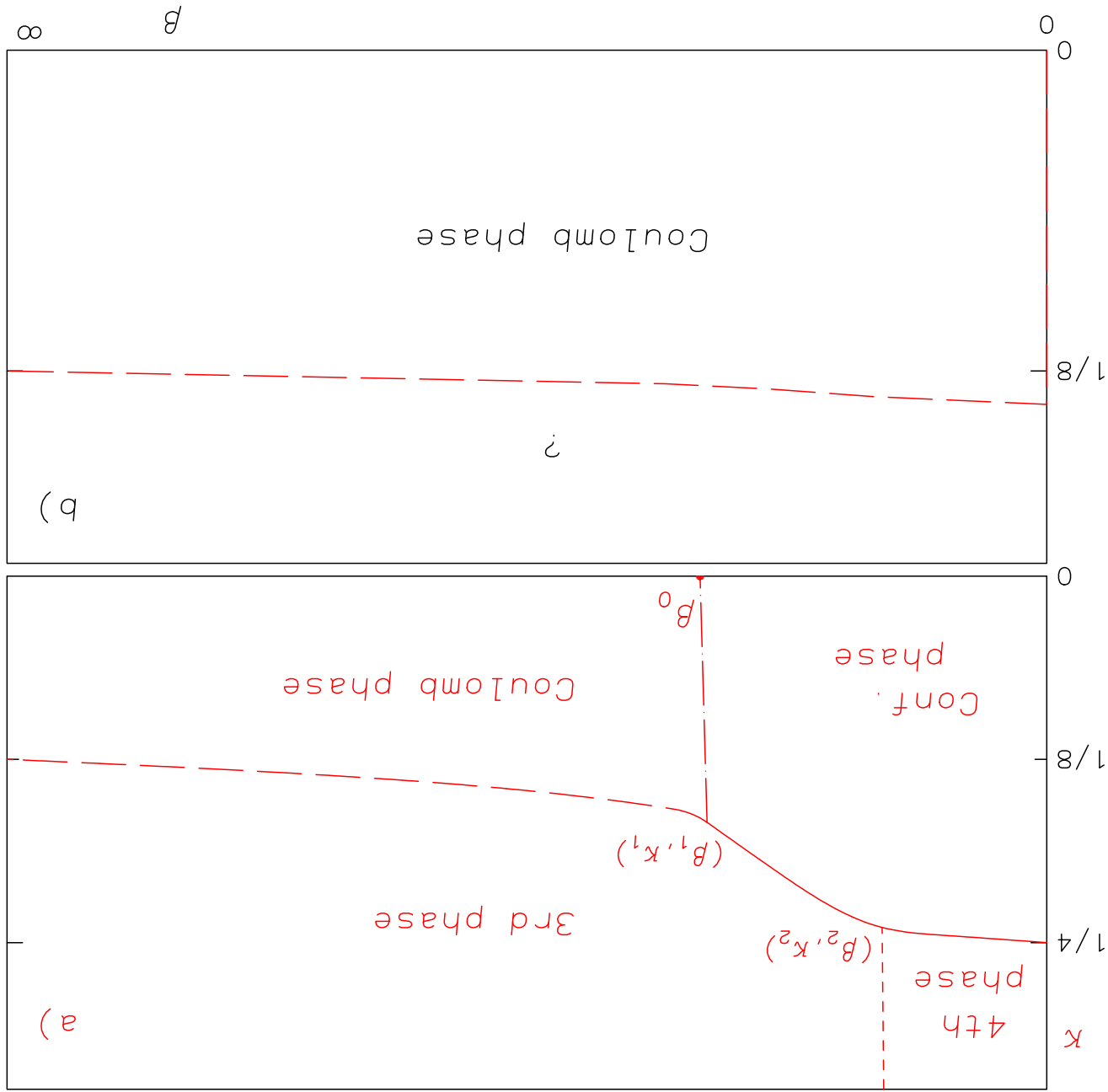
\end{document}